\documentclass[runningheads]{llncs}

\input{psfig.sty}

\begin{document}

\pagestyle{headings}

\mainmatter

\title{Synchronization in Network Structures:\\
Entangled Topology as Optimal Architecture for Network Design}

\titlerunning{Lecture Notes in Computer Science}

\author{Luca Donetti\inst{1,3}
\and Pablo I. Hurtado\inst{1,2} \and
Miguel A. Mu\~noz\inst{1}}

\authorrunning{L. Donetti, P.I. Hurtado, and M.A. Mu\~noz}

\institute{Departamento de Electromagnetismo y F\'{\i}sica de la Materia, and 
Instituto \emph{Carlos I} de F\'{\i}sica Te\'orica y Computacional \\
Facultad de Ciencias, Universidad de Granada, 18071 Granada, Spain \\
\and
Laboratoire des Collo\"ides, Verres et Nanomat\'eriaux, \\
Universit\'e Montpellier II, Montpellier 34095, CEDEX 5 France\\
\and
Departamento de Electr\'onica y Tecnolog{\'\i}a de Computadores\\
Facultad de Ciencias,
Universidad de Granada, 18071 Granada, Spain \\
\email{donetti@ugr.es, phurtado@onsager.ugr.es, mamunoz@onsager.ugr.es}
}

\maketitle

\begin{abstract}
  In these notes we study synchronizability of dynamical processes
  defined on complex networks as well as its interplay with network
  topology. Building from a recent work by Barahona and Pecora [Phys.
  Rev. Lett. {\bf 89}, 054101 (2002)], we use a simulated annealing
  algorithm to construct optimally-synchronizable networks. The
  resulting structures, known as {\it entangled networks}, are
  characterized by an extremely homogeneous and interwoven topology:
  degree, distance, and betweenness distributions are all very narrow,
  with short average distances, large loops, and small
  modularity. Entangled networks exhibit an excellent (almost optimal)
  performance with respect to other flow or connectivity properties
  such as robustness,
random walk minimal first-passage times, and good searchability. All
this converts entangled networks in a powerful concept with optimal
properties in many respects.
\end{abstract}

\section{Introduction}

It is broadly recognized that most complex systems in Nature are
organized as intricated network patterns
\cite{general,reviews}. This observation has triggered an
intense research effort aimed at understanding the organizing
principles of these networks, their structural properties, and the
interplay between topology and dynamics \cite{general,reviews}.  It
was recently recognized that the classical models of random networks
developed in graph theory were unable to describe the random but
structured, hierarchical network patterns found in Nature.  Since
then, a number of paradigmatic models (as small-world and scale-free
nets \cite{reviews}) have seen the light.  They mimic some of the
striking properties observed in real complex networks.  In any case,
network structures play an important role in many contexts ranging
from brain neural circuits, cellular function webs, ecosystems, social
networks, food webs, etc., to power grids, Internet or the world wide
web.  While most of the initial effort was put into understanding the
topological properties of networks, the interest has gradually shifted
towards the analysis of the interplay between topology and the
dynamics of network components. In general, each element (node) in a
network undergoes a dynamical process while coupled to other nodes.
The system collective behavior depends strongly on the efficiency of
communication paths, which is in turn dictated by the underlying
network topology. In this way, the network structure determines to a
large extent the possibility of a coherent response.

Complete synchronization is the most prominent example of coherent
behavior, and is a key phenomenon in systems of coupled oscillators as
those characterizing most biological networks or physiological
functions \cite{PG}. For instance, synchronized neural firing has been
suggested as specially relevant for neural signal transmission
\cite{Neural}.
From a more technological point of view, precision synchronization of
computer clocks in local area networks and the Internet is essential
for optimal network performance.
Moreover, in an
interesting twist, the dynamics toward synchronization has been
recently used as a dynamical process unveiling the underlying
community structure in complex networks \cite{Arenas}.

Here we study how synchronous behavior is affected by the network
structure. The range of stability of a synchronized state is a measure
of the system ability to yield a coherent response and to distribute
information efficiently among its elements, while a loss of stability
fosters pattern formation \cite{Pecora}.  Here we answer the following
question: {\it which is the topology that maximizes the network
synchronizability?}  \cite{DHM}.  We will construct such optimal
topologies, for any fixed number of nodes and links, by employing an
optimization procedure.  The resulting structures, that we call {\it
entangled networks}, are optimal not only for synchronizability, but
also regarding other flow or connectivity properties.

The paper is structured as follows. In section 2 we summarize the
spectral approach to synchronization, following Ref. \cite{Pecora}. In
section 3 we introduce the optimization procedure to obtain networks
with optimal synchronizability.
Section 4 discusses the relation between the emerging structures and
other optimal network designs in the literature. Finally, conclusions
and further developments are presented. A shorter presentation of this
work has been published before \cite{DHM}.

\section{Spectral Approach to Synchronization in Networks}

Consider $N$ identical oscillators at the nodes of an undirected and
unweighted graph. The state of an oscillator is represented in general
by a vector ${\bf x}_i$, $i\in [1,N]$, where $N$ is the number of
nodes. The network is characterized by its Laplacian matrix ${\bf L}$,
with elements $L_{ii}=k_i$ (the degree of node $i$), $L_{ij}=-1$ if
nodes $i$ and $j$ are connected, and $L_{ij}=0$ otherwise.  ${\bf L}$
is therefore a symmetric matrix with zero-sum rows and real,
non-negative spectrum.  The dynamics of the $i$-th node can then be
represented in a very general form as,
\begin{equation}
\frac{\textrm{d}{\bf x}_i}{\textrm{d}t} = {\bf F} ({\bf x}_i) - \sigma
\sum_{j=1}^N L_{ij} {\bf H}({\bf x}_j) \quad .
\label{ecevol}
\end{equation}
Here ${\bf F} ({\bf x})$ and ${\bf H} ({\bf x})$ are unspecified
evolution and coupling functions, respectively. In particular, ${\bf
F} ({\bf x})$ controls the dynamics of the uncoupled oscillators,
while ${\bf H} ({\bf x})$ specifies how variables at different nodes
couple together. Most dynamical processes studied in the literature
regarding synchronization can be recasted in forms equivalent to
eq. (\ref{ecevol}) (see \cite{Pecora} for more general couplings).

In the synchronized state all oscillators behave identically at all
times. That is, ${\bf x}_i (t)= {\bf x}^s(t)$ $\forall i \in [1,N]$,
where ${\bf x}^s(t)$ is solution of the uncoupled equation $\dot{\bf
x}^s = {\bf F}({\bf x}^s)$ ($\dot{\bf x}$ represents the time
derivative of ${\bf x}$).  The $N-1$ synchronization constraints ${\bf
x}_1(t) = {\bf x}_2(t) = \ldots = {\bf x}_N(t)$ define a
\emph{synchronization manifold}. This manifold is invariant owing to
the zero-sum row condition in the Laplacian matrix ${\bf L}$
\cite{Pecora}. We are interested here in the stability of the
synchronized state. For that, we introduce small perturbations ${\bf
\xi}_i$ such that ${\bf x}_i = {\bf x}^s + {\bf
\xi}_i$, and expand to first order to arrive at: $\dot{\bf \xi}_i =
\sum_{j=1}^N
\big[\partial {\bf F}({\bf x}^s) \delta_{ij} - \sigma L_{ij} \partial
{\bf H}({\bf x}^s) \big] \cdot {\bf \xi}_i$,
where $\partial {\bf M}$ stands for the Jacobian of a matrix ${\bf
M}$. Diagonalization of {\bf L} transforms these equations into a set
of $N$ independent equations for the normal modes
\cite{Pecora,DHM}:
\begin{equation}
\frac{\textrm{d}{\bf y}_k}{\textrm{d}t} = \big[\partial {\bf F}({\bf x}^s) - 
\sigma \lambda_k \partial {\bf H}({\bf x}^s) \big] \cdot {\bf y}_k \quad ,
\label{ecdiag}
\end{equation}
where $\lambda_k$, $k\in [1,N]$, are the eigenvalues of ${\bf L}$,
$0=\lambda_1 \le \lambda_2 \le \ldots \le \lambda_N$.
All the resulting equations have the same form $\dot{\bf y} =
\big[\partial {\bf F}({\bf x}^s) - \alpha \partial {\bf H}({\bf x}^s)
\big] \cdot {\bf y}$ for some positive constant $\alpha$. The
synchronized state ${\bf x}^s$ will be stable if and only if all the
perturbations fade away in time. This is equivalent to demanding the
maximum Lyapunov exponent $\eta_{max}(\alpha)$ associated with the
normal modes orthogonal to the synchronization manifold to be
negative.  The function $\eta_{max}(\alpha)$ has been called the
\emph{master stability function} in literature \cite{Pecora}, and its
dependence on $\alpha$ has an universal ``V-shape'' for most
oscillating systems. In particular, $\eta_{max}(\alpha)$ is negative
only in an interval $\alpha_A < \alpha < \alpha_B$.
The synchronized state will
be stable if all the non-trivial eigenvalues of ${\bf L}$,
$\{\lambda_k: k=2,\ldots,N\}$, lie within the interval
$[\alpha_A/\sigma,\alpha_B/\sigma]$. The following inequality then
guarantees that there always exists a coupling strength $\sigma$ for
which the synchronized state is stable,
\begin{equation}
Q\equiv \frac{\lambda_N}{\lambda_2} < \frac{\alpha_B}{\alpha_A} \quad.
\label{eigenratio}
\end{equation} 
It is important to notice that the left hand side in the above
inequality depends exclusively on the network topology, while the
right hand side depends only on the dynamics (through ${\bf x}^s$,
${\bf F}$ and ${\bf H}$). The $\sigma$ range for which the
synchronized state is stable, $\sigma \in
[\frac{\alpha_A}{\lambda_2},\frac{\alpha_B}{Q\lambda_2}]$, is larger
for smaller eigenratios $Q$. In this way, networks with very small $Q$
will exhibit very good (robust) synchronization properties for a
generic dynamics. The aim of this paper is to find and characterize
network topologies minimizing the eigenratio $Q$.

\section{Optimizing Synchronizability: Entangled Networks}

Most studies up to now have explored the value of the eigenratio $Q$
for different pre-existing network topologies found in literature, as
for instance small-world or scale-free networks, trying to identify
key topological features affecting $Q$. In this way, it has been
reported that small-worlds have smaller $Q$ than regular or purely
random graphs \cite{Pecora}, and this has been attributed to a smaller
average distance between nodes in small-worlds.  However, other works
\cite{homog} have concluded recently that $Q$ decreases as some
heterogeneity measures decrease, even if the average distance between
nodes increases in the process. On the other hand, synchronizability
is enhanced in weighted complex networks \cite{weight}.

\begin{figure}
\centerline{\psfig{figure=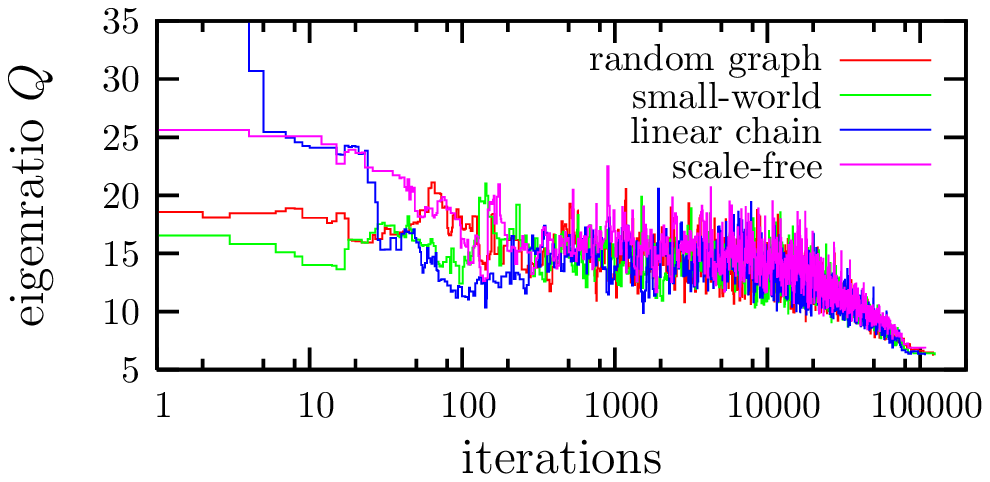,height=3.5cm}
\psfig{figure=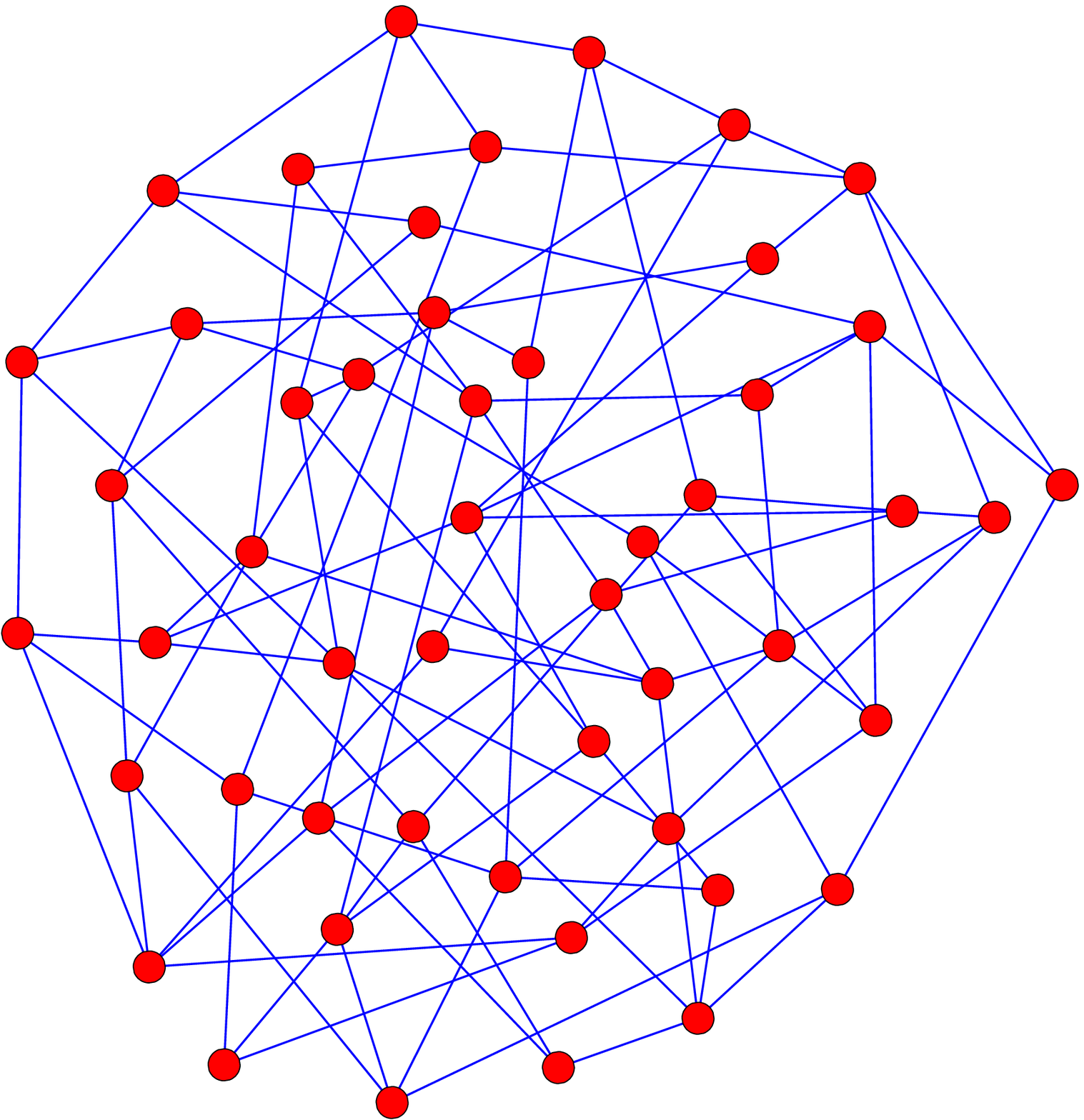,height=4cm}}
\caption{Evolution of the eigenratio $Q$ during the optimization
procedure for different initial conditions.  Here $N=50$, $\langle
k\rangle = 4$. In all cases, the algorithm converges to very
homogeneous networks as the one depicted, with very similar values of
$Q$.}
\label{fig1}
\end{figure}

In this paper we undertake a constructive approach to determine the
network topology that optimizes synchronization.  In order to do so,
we devise a modified simulated annealing algorithm \cite{Penna} to
numerically minimize $Q$. We start from graphs with $N$ nodes and a
fixed average degree $\langle k\rangle$. At each step, a new graph is
obtained by the random deletion of $m$ links and addition of $m$ new
ones, where $m$ is extracted from an exponentially decaying
distribution \cite{DHM}. The new graph is rejected if the resulting
network is disconnected; otherwise, it is accepted with probability
$p=\textrm{min}(1,[1-(1-q)\delta Q/T]^{1/(1-q)})$, where $\delta
Q=Q_{final}-Q_{initial}$ is the eigenratio change in the rewiring, and
$T$ is a temperature-like parameter. For $q \to 1$ we recover the
standard Metropolis algorithm with \emph{Hamiltonian} $Q$, while
$q=-3$ turns out to be the most efficient value (results do not depend
on the choice of the deformation parameter $q$, but convergence
times do \cite{Penna}). The first $N$ rewirings are performed at
$T=\infty$, and they are used to calculate a new $T$ such that the
largest $\delta Q$ among the first $N$ ones would be accepted with a
large probability. 
$T$ is kept fixed for $100 N$ rewiring attempts or $10 N$ accepted
ones, whichever occurs first. Then $T$ is decreased by $10\%$ and the
process is repeated until there are no more changes during five
successive temperature steps, assuming in this case that the optimal
network topology has been found.  Most of these details can be
modified without affecting the final outcome.  The major drawback in
the algorithm is that $Q$ is a global observable slow to compute. For
small enough $N$ ($\le 30$), the emerging optimal topology found is
unique, while for larger $N$ (we have optimized networks with $N$ up
to $2000$) the output may change slightly from run to run, meaning
that the eigenratio absolute minimum is not always reached due to the
presence of metastable states. Nevertheless, the final values of $Q$
are very similar for different runs (see Fig. \ref{fig1}), meaning
that a reasonably good approximation to the optimal topology is always
found \cite{DHM}.

\begin{figure}
\centerline{\psfig{figure=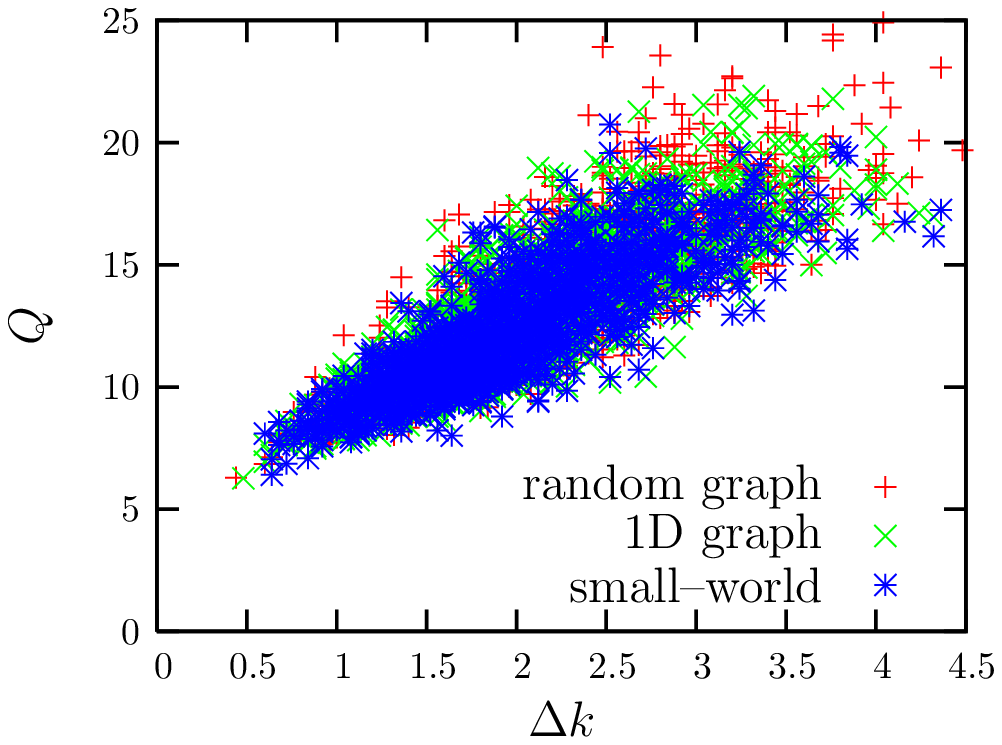,height=3.8cm}
       \psfig{figure=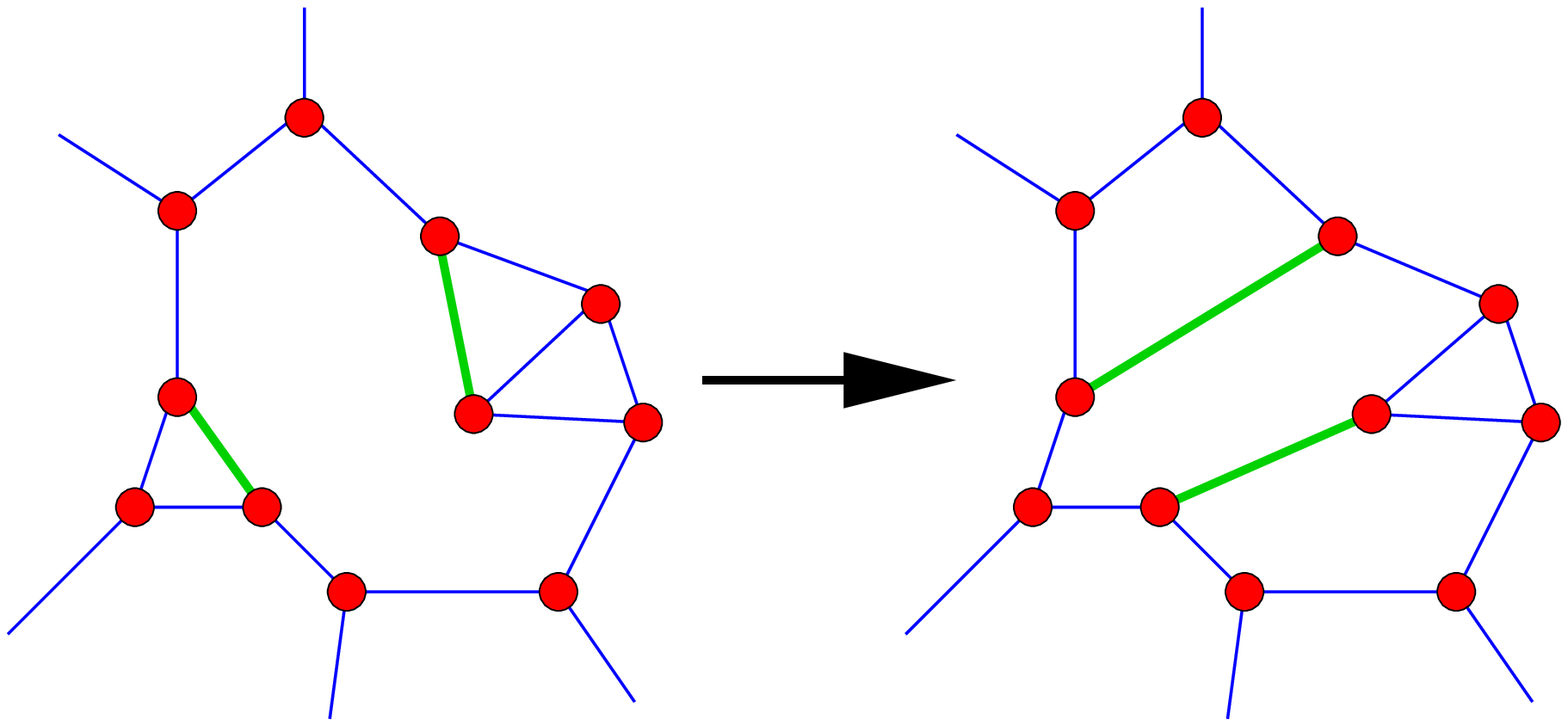,angle=-90,height=4.0cm}
       \psfig{figure=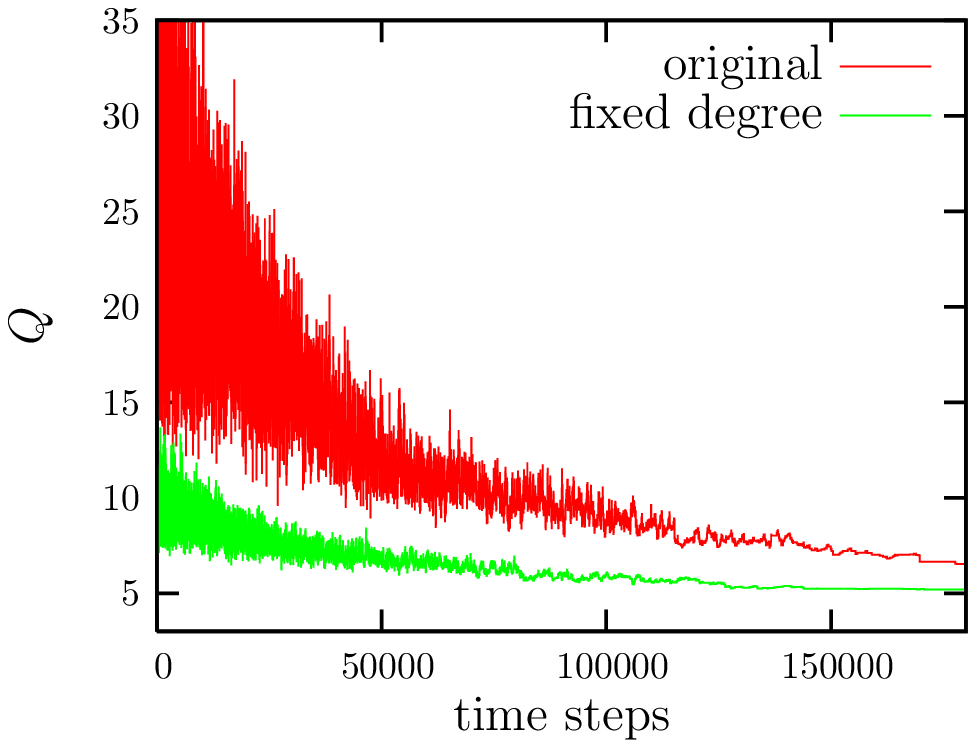,height=3.8cm}}
\caption{Left: $Q$ vs standard deviation of the degree distribution
for $N=50$, $\langle k\rangle =4$ and three different initial
conditions. Center: sketch of pair rewiring trial. Right: Eigenratio
$Q$ vs algorithmic steps for both minimization procedures (see text).}
\label{fig2}
\end{figure}

We measure different topological observables during the minimization
process to unveil the main traits of the emerging structures. In
simple terms, we observe that as $Q$ decreases the network becomes
more and more homogeneous. This means that the standard deviation of
distributions of most topological observables
decreases as $Q$ decreases. This is true in particular for the node
degree distribution, see Fig. \ref{fig2}.a. We have used this degree
homogeneity to improve the efficiency our optimization procedure by
initializing the algorithm with regular networks (i.e. all nodes with
the same degree), and restricting the rewiring steps to changes that
leave the degree of each node unchanged (by randomly selecting
\emph{pairs} of links and exchanging their endpoints; see Fig.
\ref{fig2}.b). The resulting algorithm converges much faster to the
optimal network, and yields lower final eigenratios $Q$ when the
original one get trapped in a metastable state (Fig~\ref{fig2}.c).

\begin{figure}[t]
\centerline{\psfig{figure=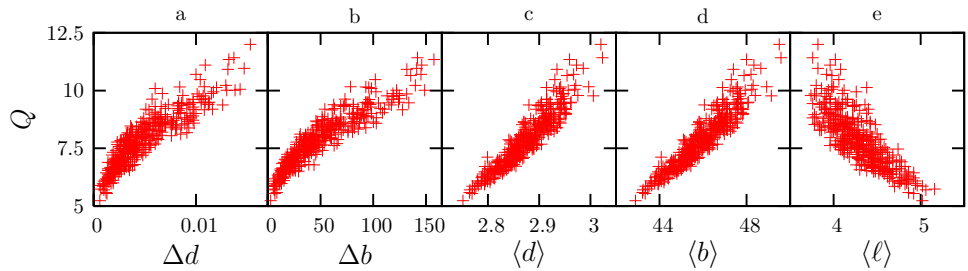,height=3.5cm}}
\caption{Standard deviations of the node distance distribution (a),
standard deviation of the betweenness distribution (b), average
distance (c), average betweenness (d), and shortest-loop average
length (e) as a function of the eigenratio $Q$.}
\label{fig3}
\end{figure}

In Figs. \ref{fig3}.a-b we show the standard deviation of the average
node-to-node distance and average betweenness, respectively, versus
$Q$ during an optimization run started from a random regular graph. Both
observables exhibit the aforementioned tendency towards
homogeneity. Particularly remarkable is the narrow betweenness
distribution (Fig. \ref{fig3}.b), which is in marked contrast with the
broad betweenness distributions observed in networks with strong
community structure \cite{communities}.
In addition, the averaged distance and betweenness also tend to
decrease with $Q$, though they are less sensitive than their
corresponding standard deviations, see Figs. \ref{fig3}.c-d. Another
key feature of the optimal structures is the absence of short
loops. This can be characterized by the \emph{girth} (length of the
shortest loop) or, better, via the average size of the shortest loop
passing through each node. This last magnitude is shown in
Fig. \ref{fig3}.e, where it is evident that the optimal network has
very large average shortest loops. In particular, the clustering
coefficient is zero for the optimal nets since no triangles are
present.

In general, we call the emerging optimal structures \emph{entangled
networks}: all sites are very much alike (strong homogeneity) and the
links form a very intricated or interwoven structure (lack of
communities, poor modularity, and large shortest loops). Every single
site is close to any other one (short average distances) owing not to
the existence of intermediate highly connected hubs (as in scale free
networks), but as a result of a very ``democratic'' or entangled
structure in which properties such as site-to-site distance,
betweenness, and minimum-loop size are very homogeneously distributed
(see Figs. \ref{fig1}.b, \ref{fig3}.a-b).
Sharp distributions are also typical of random graph, where randomness
alone produces a statistical homogeneity. However in this case, a much
stronger homogeneity is produced during the optimization of $Q$, as
Figures~\ref{fig3}.a-b show.

\section{Relation to Other Optimal Topologies}


A natural question concerns the relation between entangled networks
and other optimal architectures found in the literature.  For
instance, recent work \cite{2peak} has focused on the optimization of
network robustness against random and/or intentional removal of nodes
(attacks). For random graphs in the large-$N$ limit, it is concluded
that the most robust networks are obtained when the degree distribution
only has a few peaks. In particular, random $k$-regular graphs
turn out to be the global optimal solution against both errors and
attacks in the robustness-optimization problem \cite{2peak}. In this
case, the error ($f_r$) and attack ($f_a$) percolation thresholds
coincide, $f_r=f_a\equiv f_c(N,k)$, with
$f_c(N,k)<f_c(\infty,k)=(k-2)/(k-1)$.  A natural question now is
whether further $Q-$minimization of these random regular graphs has
some effect on the network robustness. As shown in \cite{DHM}
the minimization of $Q$ improves significantly the network robustness,
confirming that entangled networks are optimal from the robustness
point of view. This is because entangled topologies include
correlations, absent in random networks, which enhance their
resilience. In addition, there is also evidence that networks with
properties similar to those of entangled graphs
maximize reliability against link removal \cite{DHM}.

Different models of traffic flows on complex network have been recently
studied \cite{tadic,search}. In principle highly inhomogeneous
scale-free networks perform well when the traffic is low; hubs can
provide fast transition times, while they easily jam when the traffic
increases.
With the model of \cite{search} it has been shown that if the density
of traveling information packets is above a given threshold, the
optimal network topology is a highly homogeneous, isotropic 
configuration, strongly resembling entangled graphs.
In a similar way, it has been recently reported \cite{selection} that
the interplay between network growth processes and evolutionary
selection rules gives rise in some cases to very homogeneous
structures with large minimal-loops that strongly resemble entangled
networks (see Fig. 3.c in \cite{selection}).


Also, during our optimization procedure, $\lambda_N$ is observed to
change very little with respect to $\lambda_2$, and therefore,
minimizing $Q$ is equivalent for all practical purposes to maximizing
$\lambda_2$. This provides another interesting connection with graph
theory, where it is known that regular graphs with a large $\lambda_2$
(i.e.{\it large spectral gap}), are good {\it expanders} (see
\cite{expanders,DHM} for a definition and applications). Expander graphs are
very important in computer science for various applications (as the
design of efficient communication networks or construction of
error-correcting codes) and can be proved to exhibit a rapid decay of
random-walk distributions towards their stationary state
\cite{Lovasz}. This converts entangled graphs in (almost) optimal 
for many information flow processes.

\section{Summary and Outlook}

We have introduced the concept of ``entangled networks'' \cite{DHM}.
These are constructed using an optimization principle by imposing the
eigenvalues of the Laplacian matrix to have a quotient $\lambda_N
/\lambda_2$ as small as possible, guaranteeing in this way a robust
synchronizability and coherent behavior.
The emerging topologies are extremely homogeneous: all nodes look very
much alike (constituting a topology radically distinct from scale free
networks).  Also, the node-to-node average distance tends to be small
while the average shortest loops are very large, and there is no
modular (or community) structure.  Entangled networks exhibit optimal
synchronization properties, but they are also optimal or
almost-optimal for other communication or flow properties:
robustness and resilience against errors and attacks, traffic flow in
the presence of congestion, relaxation
properties of random walks, etc. These connections make of entangled
networks a key tool in the context of complex networks.

An interesting issue concerns the existence of entangled networks in
Nature. Their construction requires a global optimization process
which is unlikely to occur in natural evolving systems. Presently, we
are working on the identification of local evolutionary rules which
give rise to locally-optimal synchronizable network patterns, or other
feasible approximations to entangled networks.

\section*{Acknowledgments}
We thank D. Cassi and P.L. Krapivsky for useful discussions, and
B. Tadi\'c and S. Thurner for inviting us to the ICCS06. Financial
support from the Spanish MCyT under project No. FIS2005-00791, EU
COSIN-project-IST2001-33555, and EU HPRN-CT-2002-00307 (DYGLAGEMEM)
are also acknowledged.

\end{document}